\documentclass[superscriptaddress,showkeys,twocolumn,nofootinbib,longbibliography]{revtex4-1}

\usepackage{amsmath}
\usepackage{amssymb}
\usepackage{graphicx}
\usepackage{epsf}
\usepackage{slashed}
\usepackage{enumitem}
\usepackage[czech,english]{babel}
\usepackage[cp1250]{inputenc}
\usepackage{hyperref}
\usepackage{xcolor}

\usepackage[only,llbracket,rrbracket,llparenthesis,rrparenthesis]{stmaryrd} 
\usepackage{accsupp} 

\newcommand{\Tr}{\mathop{\rm Tr}\nolimits}
\newcommand{\I}{\ensuremath{\mathrm{i}}}
\newcommand{\e}{\ensuremath{\mathrm{e}}}
\renewcommand{\d}{\ensuremath{\mathrm{d}}}
\newcommand{\qm}[1]{``#1''} 

\newcommand{\sgn}{\mathop{\rm sgn}\nolimits}

\usepackage{bbm} 
\newcommand{\unitmatrix}[1]{\ensuremath{\mathbbm{1}_{#1}}}

\usepackage[nodayofweek]{datetime}
\newdateformat{mydate}{\twodigit{\THEDAY}{ }\shortmonthname[\THEMONTH], \THEYEAR}
\usepackage[normalem]{ulem}

\begin{document}

\title{BPS Cho--Maison monopole}

\author{Filip Blaschke}
\email{filip.blaschke@fpf.slu.cz}
\affiliation{Faculty of Philosophy and Science, Silesian University in Opava,\\Bezru\v{c}ovo n\'am\v{e}st\'i~1150/13, 746~01 Opava, Czech Republic}
\affiliation{Institute of Experimental and Applied Physics, Czech Technical University in Prague,\\Horsk\'a 3a/22, 128~00 Praha 2, Czech Republic}

\author{Petr Bene\v{s}}
\email{petr.benes@utef.cvut.cz}
\affiliation{Institute of Experimental and Applied Physics, Czech Technical University in Prague,\\Horsk\'a 3a/22, 128~00 Praha 2, Czech Republic}

\begin{abstract}
We present exact solutions to Cho--Maison magnetic monopole in a family of effective electroweak models that have a Bogomol'nyi--Prasad--Sommerfield (BPS) limit. We find that the lower bound to the mass of the magnetic monopole is $M \geq 2\pi v/ g \approx 2.37\,\,\mathrm{TeV}$. We argue that this bound holds universally, not just in theories with a BPS limit.
\end{abstract}


\maketitle


\section{Introduction}

Dirac's monopole \cite{Dirac}, later generalized to the dyon by Schwinger \cite{Schwinger}, remains the most fruitful theoretical idea that is yet to be experimentally verified to date. Dirac's famous quantization condition linking together the magnitude of the electric charge $e$ and the magnetic charge $q = 2\pi n/e$ is but the first of its many interesting consequences. 

Originally, Dirac imagined his monopole as an optional component of (quantum) electrodynamics. However, the idea of a magnetic monopole quickly pollinated all major disciplines of theoretical physics. It was introduced into non-Abelian $SU(2)$ gauge theory by Wu and Yang \cite{Wu}. Later, 't~Hooft \cite{tHooft} and Polyakov \cite{Polyakov:1974ek} independently showed that a non-singular $SU(2)$ monopole configuration can be achieved with adjoint scalar fields. 

Subsequently, Bogomol'nyi \cite{Bog} and, independently, Prasad and Sommerfield \cite{Prasad:1975kr} discovered that in the limit of the vanishing potential the monopole can be found in an analytic form. This limit, henceforth known as the Bogomol'nyi--Prasad--Sommerfield (BPS) limit, became an important tool in both finding and studying classical solutions in gauge theories coupled with scalar matter. Among other things, in the BPS limit, one can directly calculate the mass without the need for integration. Furthermore, the BPS limit allows the construction of multi-particle solutions as static configurations since, in this limit, the repulsive force mediated by gauge fields is precisely balanced by the attractive force mediated by scalars. This also enables an approximate analysis of scattering without the need to solve the underlying partial differential equations \cite{Manton}. In short, the BPS limit provides us with a useful analytical window for studying its non-BPS (and more realistic) counterparts.

The electroweak model of Weinberg \cite{Weinberg:1967tq} and Salam \cite{Salam:1968rm} was long thought to be void of monopoles. The standard argument was that the underlying quotient space $SU(2)_L \times U(1)_Y/U(1)_{\rm em}$ has only a trivial second homotopy group and, hence, cannot support a monopole. While this is true, Cho and Maison \cite{Cho:1996qd} showed that the desired topology can be found elsewhere, namely in the (normalized) Higgs doublet, which can be regarded  as a $\mathbb{C}\mathrm{P}^1$ coordinate. This gives us a second homotopy $\pi_2\bigl(\mathbb{C}\mathrm{P}^1\bigr) = \mathbb{Z}$. In addition, the Cho--Maison monopole---like Dirac's monopole---has a point singularity in the magnetic field and a line singularity in $U(1)_Y$ gauge fields. As such, the Cho--Maison monopole represents a peculiar mixture of the ideas of 't~Hooft--Polyakov and Dirac.

With experiments such as MoEDAL \cite{moedal} currently searching for magnetic monopoles, the problem of estimating the mass of the Cho--Maison  monopoles is especially urgent. The issue is that the Cho--Maison monopole has a divergent energy and one needs a sufficiently robust regularization method to obtain a valid estimate. In their paper \cite{Cho:2013vba}, Cho, Kim, and Yoon (CKY) showed that by modifying the hypercharge coupling constant to depend on the magnitude of the Higgs doublet, i.e.,
\begin{eqnarray}
\label{eq:term}
{\mathcal L} &\supset& -\frac{1}{4}\epsilon\bigg(\frac{|H|}{v}\bigg)B_{\mu\nu}B^{\mu\nu}
\,,
\end{eqnarray} 
where $\epsilon(1)=1$ becomes unity in the Higgs vacuum $|H|\to v$, it is possible to have finite Cho--Maison monopoles within the range 4 to 10 TeV. For concreteness, they showed that for $\epsilon \sim |H|^8$ the monopole mass is $(4\pi/e^2) M_W \approx 7.2\,\,\mathrm{TeV}$. This was picked up by Ellis et al.~\cite{Ellis:2016glu} who pointed out that such a choice leads to an unrealistically high $H\gamma\gamma$ vertex and proposed a whole series of alternative forms for $\epsilon$. Using the principle of maximum entropy to determine unconstrained parameters in $\epsilon$, they showed that some choices can get as low as $\approx 5.5\,\,\mathrm{TeV}$, opening up the possibility of pair-production of Cho--Maison monopoles at LHC.

In this paper, we continue this story by providing a definitive \emph{lower} bound on the mass of the Cho--Maison monopole, which is $M \geq 2\pi v/ g \approx 2.37\,\,\mathrm{TeV}$.  We achieve this by constructing a family of effective electroweak models that has a BPS limit. In this limit, the mass of the Cho--Maison monopole can be found analytically and it is determined purely by the asymptotic behavior of fields. We then show that CKY theory can be reformulated as a BPS theory with additional positive contributions. This demonstrates that the monopoles in the CKY model cannot be lighter (and in fact must be heavier) than the BPS monopoles presented in this paper.
 
A further advantage of considering a BPS extension of the CKY model lies in the fact that it motivates non-canonical modifications of the theory. While in the CKY model the introduction of the $\epsilon$ term is rather ad hoc, in this work, it would not be possible to achieve the BPS limit without it. As it turns out, we actually need modifications to the $SU(2)_L$ kinetic term as well.

The idea to make gauge couplings of the theory into functions of gauge invariants is rather natural from the point of view of the effective Lagrangian approach. These terms can be understood as a finite or infinite sum of loop corrections to the  electroweak theory or some other theory beyond the SM. Even though the top-down justification for their precise structure would be difficult and beyond the scope of this paper, in general, there could be many different dynamics that cause them.
 
Moreover, the idea of field-dependent gauge coupling of the type \eqref{eq:term} arises quite naturally within the context of supersymmetric theories but its utility reaches far beyond. For instance, this concept has been successfully used in localizing zero modes of non-Abelian gauge fields on the domain wall \cite{Ohta:2010fu}. The physical reasoning is that such a term classically mimics the effects of confining vacuum outside of the domain wall, which acts as a dual superconductor forcing the field lines to be squeezed inside the domain wall. Recently, a similar idea was used in constructing a 5D $SU(5)$ GUT scenario in the background of five domain walls \cite{Arai:2017lfv}. There, the gauge-kinetic scalar term was responsible for dynamical symmetry breaking of $SU(5)$ to the SM gauge group  \cite{Arai:2017ntb} via a non-coincident configuration of walls, similar to D-branes. In this sense, we view the modifications of the form \eqref{eq:term} as yet another application of the position-dependent gauge coupling idea.
 

The paper is organized as follows. In Sec.~\ref{sec:II}, we introduce our model and derive BPS equations. In Sec.~\ref{sec:III}, we solve the BPS equations in the spherically symmetric case and present a general formula for monopole mass and charge. We also discuss the properties of monopoles in various special cases. Sec.~\ref{sec:IV} is devoted to the general BPS theory, which is used in Sec.~\ref{sec:V} to demonstrate the universality of the lower bound on the monopole mass. Lastly, we discuss our results in Sec.~\ref{sec:VI}.

\raggedbottom

\section{Model and the BPS equations}
\label{sec:II}

Let us consider the following modification of the bosonic part of the electroweak model:
\begin{eqnarray}
\label{eq:model}
{\mathcal L}_{\rm eff} &=& 
-\frac{v^2}{4 g^2 |H|^2}\Tr\big[F_{\mu\nu}F^{\mu\nu}\big]
\nonumber \\ && {}
+ \frac{v^2}{2g^2 |H|^6}\Tr\big[F_{\mu\nu} H H^{\dagger}\big]\Tr\big[F^{\mu\nu} H H^{\dagger}\big]
\nonumber \\ && {}
-\frac{1}{4g^{\prime 2}}f^{\prime 2}\bigg(\frac{\sqrt{2}|H|}{v}\bigg)B_{\mu\nu}B^{\mu\nu} 
\nonumber \\ && {}
+|D_\mu H|^2-\frac{\lambda}{2}\bigg(H^{\dagger}H -\frac{v^2}{2}\bigg)^2
\,,
\end{eqnarray}
where $v = 246 \, \mathrm{GeV}$ is the electroweak scale. The field strength tensors of the $SU(2)_L$ and $U(1)_Y$ gauge groups are
\begin{eqnarray}
F_{\mu\nu} &=& \bigl(\partial_\mu A_\nu^a-\partial_\nu A_\mu^a -\varepsilon^{abc}A_{\mu}^bA_{\nu}^c\bigr)\tau^a \,, \\
B_{\mu\nu} &=& \partial_\mu B_\nu -\partial_\nu B_\mu \,,
\end{eqnarray}
and the covariant derivative of the Higgs doublet $H$ is
\begin{eqnarray}
D_\mu H &=& \bigg(\partial_\mu \unitmatrix{2} + \I A_\mu^a \tau^a + \I \frac{1}{2}  B_\mu \unitmatrix{2} \bigg)H \,.
\end{eqnarray}
The $SU(2)_L$ generators $\tau^a$ satisfy $[\tau^a,\tau^b] = \I \varepsilon^{abc} \tau^c$ and are normalized as $\Tr [\tau^a\tau^b] = \frac{1}{2}\delta^{ab}$.

In the model \eqref{eq:model}, we introduced a function $f$, whose derivative squared modifies the kinetic term of the $U(1)_Y$ gauge field in the same manner as in Refs.~\cite{Cho:2013vba, Ellis:2016glu}\footnote{The function $f^{\prime 2}/g^\prime$ corresponds to the function $\epsilon$ in the notation of \cite{Ellis:2016glu}; compare Eq.~\eqref{eq:term}.}. As shown in \cite{Cho:2013vba}, the purpose of this function is to modify the permittivity of the $U(1)_Y$ gauge field and, consequently, make the energy of the monopole solution finite. Requiring that in the vacuum ($|H| = v/\sqrt{2}$) the kinetic term is unmodified, we adopt the normalization
\begin{eqnarray}
\label{fprimein1}
f^{\prime 2}(1) &=& 1 \,.
\end{eqnarray}
Notice that this condition does not impact the generality of our discussion, since any  $f^{\prime 2}(1)\not = 1$ can be absorbed into the definition of $g^\prime$.

In contrast to \cite{Ellis:2016glu} or \cite{Cho:2013vba}, however, we also modify the $SU(2)_L$ gauge structure of the theory, as is shown in the first two lines of Eq.~\eqref{eq:model}. The motivation for these two effective terms is to make the theory possess the BPS limit. Interestingly, while we have virtually absolute freedom in modifying the effective permittivity of $U(1)_Y$ embodied in the function $f^\prime$, we find that modifications of $SU(2)_L$ gauge terms are completely fixed by the requirement that the mixed term
\begin{eqnarray}
&\sim& B_{\mu\nu}\Tr\big[F^{\mu\nu} H H^{\dagger}\big]
\end{eqnarray}
disappear. The main benefit of not having such a term 
is that the BPS equations are simplest and easiest to solve. The price we pay, however, is the apparent pole at $|H|=0$. Let us stress that, despite appearances, the model \eqref{eq:model} is well defined around the Higgs vacuum (if not identical to the electroweak model), as we will see later in this section. In fact, it is best not to perceive Eq.~\eqref{eq:model} as a model for phenomenological inquiry; rather, it is a theoretical laboratory, where Cho--Maison monopoles are incarnated in their analytically simplest form. 

In Sec.~\ref{sec:IV}, we drop this \qm{anti-mix term} policy and explore the Ivory Tower of all possible BPS theories to its full heights. We find a much richer family of BPS theories, where one can completely avoid the above-mentioned singular appearance. Surprisingly, we find that the BPS mass of the Cho--Maison monopole has the same form in any BPS theory. With this in mind, we consider the results presented in this section to be representative and generic. 


Notice that the model \eqref{eq:model} is singular in the symmetric limit $v\to 0$ and, therefore, UV incomplete. While this may be troubling for the theory at the quantum level, for our purely classical purposes this does not constitute a serious issue and it is outside the scope of this paper.

Let us now show that for the model \eqref{eq:model} the BPS limit exists. It proves convenient to decompose the Higgs doublet $H$ as
\begin{eqnarray}
\label{HCP1}
H &=& \frac{1}{\sqrt{2}} v \rho \, \xi \,, 
\end{eqnarray}
where the $\mathbb{C}\mathrm{P}^1$ field $\xi$ (a doublet) is normalized as $\xi^{\dagger}\xi = 1$ and the radial field $\rho$ (a singlet) is real. First, let us write the energy density for static configurations (i.e., $\partial_0 = A_0^a = B_0 = 0$):
\begin{eqnarray}
{\mathcal E} &=& 
\frac{2}{g^2 \rho^2} \Big|\Big(M_i -\xi^{\dagger} M_i  \xi\Big)\xi\Big|^2 + \frac{1}{2g^{\prime 2}}f^{\prime 2}\bigl(\rho\bigr)G_i^2
\nonumber \\ && {}
+ |D_i H|^2 +\frac{\lambda v^4}{8}\Bigl(\rho^2 -1\Bigr)^2 \,,
\end{eqnarray}
where we introduce the $SU(2)_L$ and $U(1)_Y$ \qm{magnetic} fields
\begin{eqnarray}
M_i &=& 
\frac{1}{2}\varepsilon_{ijk}F_{jk} = 
\varepsilon_{ijk}\bigg(\partial_j A_k^a-\frac{1}{2}\varepsilon^{abc}A_j^bA_k^c\bigg)\tau^a \,, 
\\
G_i &=& 
\frac{1}{2} \varepsilon_{ijk} B_{jk}
= 
\varepsilon_{ijk}\partial_j B_k
\end{eqnarray} 
and where we use the identity
\begin{equation}
\Tr\bigl[M_i^2\bigr]-2 \big(\xi^{\dagger} M_i \xi\big)^2 
 = 2\big|\big(M_i -\xi^{\dagger} M_i  \xi\big)\xi\big|^2 \,,
\end{equation}
so that the positive definiteness of $\mathcal{E}$ is manifest.

In the limit $\lambda \to 0$ we can complete the energy density into a perfect square plus a total derivative:
\begin{eqnarray}
{\mathcal E} &=&
\bigg|D_i H - \eta\frac{\sqrt{2}}{g\rho} \Big(M_i -\xi^{\dagger} M_i  \xi\Big)\xi -\tilde\eta\frac{1}{\sqrt{2}g^{\prime}}f^{\prime}(\rho)G_i\, \xi\bigg|^2 
\nonumber \\ &&  {} 
+\eta\frac{v}{g}\partial_i \Big(\xi^{\dagger} M_i \xi\Big)  + \tilde\eta\frac{v}{g^{\prime}}\partial_i\Big(f(\rho)G_i\Big) 
\,,
\end{eqnarray}
where $\eta = \pm 1$ and $\tilde\eta = \pm 1$ are some signs and where we use the Bianchi identities $\partial_i M_i^a -\varepsilon^{abc} A_i^{b}M_i^c= 0$ and $\partial_i G_i = 0$.\footnote{However, for a configuration of monopoles $\partial_i G_i \neq 0$; in that case, strictly speaking, we should introduce into the Lagrangian \eqref{eq:model} a source term for the monopoles $\sim j_\mu B^{\mu}$. Alternatively, we can say that our model is not based on $\mathbb{R}^{3}$ but rather on the space with the origin removed $\mathbb{R}^{3}/\{0\}$ (for a single monopole), where $\partial_i G_i = 0$. In this paper, we will adopt this somewhat sloppy but convenient attitude as it does not change the results, as far as static configurations are concerned.} Thus, we obtain the Bogomol'nyi bound
\begin{eqnarray}
\label{Bogomolnybound}
{\mathcal E} &\geq&
\eta\frac{v}{g}\partial_i \Big(\xi^{\dagger} M_i \xi\Big)  + \tilde\eta\frac{v}{g^{\prime}}\partial_i\Big(f(\rho)G_i\Big) 
\,,
\end{eqnarray}
which is saturated if the BPS equations
\begin{eqnarray}
\label{eq:bps}
D_i H &=&
\eta\frac{\sqrt{2}}{g\rho} \Big(M_i -\xi^{\dagger} M_i  \xi\Big)\xi +\tilde\eta\frac{1}{\sqrt{2}g^{\prime}}f^{\prime}(\rho)G_i\, \xi
\hspace{1cm}
\end{eqnarray}
are satisfied. Field configurations that solve BPS equations are generically called BPS solitons and are necessarily solutions of the equations of motion.

Let us now multiply the BPS equations \eqref{eq:bps} from the left by $\xi^\dag$ to obtain
\begin{equation}\label{eq:bpss}
v \rho \Big( \xi^\dag D_i\xi\Big) + v \Bigl( \partial_i \rho + \frac{\I}{2} B_i \rho \Bigr) = \frac{\tilde\eta}{g^{\prime}}f^{\prime}(\rho)G_i \,,
\end{equation}
where 
\begin{eqnarray}
D_i \xi &=& \big(\partial_i \unitmatrix{2} + \I A_i^a \tau^a \big) \xi  \,.
\end{eqnarray}
If we now subtract from Eq.~\eqref{eq:bpss} its Hermitian conjugate we obtain, after trivial rearranging of the terms,
\begin{eqnarray}
\label{eq:ident}
B_i &=& 2 \I \xi^\dag D_i \xi \,,
\end{eqnarray}
i.e., we obtain an expression for the $U(1)_Y$ gauge field $B_i$ in terms of $\xi$ and $A_i^a$. 
On the other hand, summing Eq.~\eqref{eq:bpss} with its Hermitian conjugate yields
\begin{eqnarray}
\label{eq:rho1}
\frac{\partial_i \rho}{f^\prime(\rho)} &=&  \frac{\tilde \eta}{vg^\prime} G_i \,.
\end{eqnarray}
Lastly, using the expressions for $B_i$ and $\partial_i\rho$, we can rewrite the BPS equations \eqref{eq:bps} into the reduced form
\begin{eqnarray}
\label{eq:bpsred}
\bigl(\unitmatrix{2} -\xi\xi^{\dagger}\bigr)\bigg(D_i \xi - \eta\frac{2}{gv\rho^2} M_i\xi \bigg) 
&=& 0 \,.
\end{eqnarray}

\section{BPS Cho--Maison monopole}
\label{sec:III}

\subsection{Spherically symmetric Ansatz}

Let us now solve the BPS equation \eqref{eq:bps} for a spherically symmetric monopole. We employ the hedgehog Ansatz for $SU(2)_L$ gauge fields:
\begin{eqnarray}
\label{hedgehog}
A_i^a\tau^a &=& \big(1-K(r)\big)\varepsilon_{ijk}\frac{x_k \tau^j}{r^2} \,,
\end{eqnarray}
where $K(r)$ is some function of the radial coordinate. Furthermore we set
\begin{eqnarray}
\label{eq:formxi}
\xi &=& \I
\left(\begin{array}{c} \sin(\theta/2) \, \e^{-\I\varphi} \\ -\cos(\theta/2) \end{array}\right)
\,,
\end{eqnarray}
where $\varphi$, $\theta$ are respectively the azimuthal and polar angles. 

The hedgehog Ansatz \eqref{hedgehog} leads to
\begin{eqnarray}
M_i &=& \frac{1}{r^2}\bigg[ \frac{x_ix_a}{r^2} (1-K^2) -\bigg(\delta_{ia} - \frac{x_ix_a}{r^2}\bigg)rK^\prime \bigg]\tau^a \,.
\hspace{1cm}
\end{eqnarray}
Furthermore, it holds that
\begin{eqnarray}
\xi^{\dagger}\tau^a \xi &=& -\frac{x^a}{2r} \,,
\end{eqnarray}
which together with Eq.~\eqref{hedgehog} implies $A_i^a\xi^{\dagger}\tau^a\xi = 0$. Using the formula \eqref{eq:ident} for $B_i$ we obtain\footnote{From the second expression for the connection $B_i$, it is particularly easy to see that it is singular along the negative $z$-axis. It can be shown that the direction and even the shape of this Dirac string can be arbitrarily changed by gauge transformations and hence represents an unphysical object.}
\begin{equation}
B_i = 2i \xi^{\dagger}\partial_i \xi = (1-\cos\theta) \partial_i\varphi \ = \ -\varepsilon_{ij3} \partial_j \log(r+z)
\label{Bi}
\end{equation}
and correspondingly
\begin{eqnarray}
G_i &=& \frac{x_i}{r^3} \ = \ -\partial_i \frac{1}{r} \,.
\end{eqnarray}

Consequently, under the spherically symmetric Ansatz \eqref{hedgehog} and \eqref{eq:formxi} the BPS equations \eqref{eq:rho1} and \eqref{eq:bpsred} reduce to
\begin{subequations}
\label{EOMspherical}
\begin{eqnarray}
\label{eq:rho}
\frac{\rho^{\prime}}{f^{\prime}(\rho)} &=& \frac{\tilde \eta}{v g^{\prime} r^2} \,,
\\
\label{eq:K}
\frac{K^{\prime}}{K} &=& \eta\frac{g v}{2}\rho^2 \,,
\end{eqnarray}
\end{subequations}
which are easily solved as
\begin{subequations}
\label{solutions_general}
\begin{eqnarray}
\label{solutions_general_rho}
\rho(r) &=& F^{-1}\bigg[F\big(\rho(\infty)\big) - \tilde \eta\frac{1}{g^\prime v r} \bigg] \,,
\\
\label{solutions_general_K}
K(r) &=& K(0) \exp\bigg\{\eta\frac{g v}{2}\int\limits_{0}^{r}\d r^{\prime}\,\rho^2(r^{\prime})\bigg\} \,,
\end{eqnarray}
\end{subequations}
where 
\begin{eqnarray}
F(\rho) &\equiv& \int^{\rho}\!\frac{\d\rho^\prime}{f^\prime(\rho^\prime)} \,.
\end{eqnarray}

The boundary condition is $\rho(\infty) = 1$ so that the Higgs field acquires a proper vacuum expectation value at spatial infinity. For $K(r)$, we set $K(0) = 1$ in order for the $SU(2)_L$ gauge fields \eqref{hedgehog} to be regular at the origin. Similarly, regularity of the Higgs field at the origin requires $\rho(0) = 0$. 

The energy density (given by the Bogomol'nyi bound \eqref{Bogomolnybound}) can be, using Eqs.~\eqref{EOMspherical}, rewritten as
\begin{eqnarray}
\label{eq:enKform}
\mathcal{E} &=& \frac{v^2}{2}\frac{\rho^2K^2}{r^2}  + \frac{1}{g^{\prime 2}} \frac{f^{\prime 2}(\rho)}{r^4} \,.
\end{eqnarray}
To make the total energy (mass) integral $M = 4\pi \int_{0}^{\infty} \d r \, r^2 \mathcal{E}$ converge, we need to have $K(\infty) = 0$, which implies
\begin{eqnarray}
\label{etasign}
\eta &=& -1 \,,
\end{eqnarray}
while the convergence at the lower limit requires that $f^\prime(\rho(0)) = f^\prime(0) = 0$.

To summarize, we have the following initial and boundary conditions on $\rho(r)$ and $K(r)$ and on the function $f^\prime(\rho)$:
\begin{subequations}
\label{conditions}
\begin{align}
K(0)     &=1 \,, & \rho(0)     &=0 \,, & f^\prime(0)&=0 \,, \\
K(\infty)&=0 \,, & \rho(\infty)&=1 \,, & \big|f^\prime(1)\big|&=1 \,.
\end{align}
\end{subequations}

\subsection{The mass and its lower bound}

Recall that in the BPS limit the energy density is given as a total derivative:
\begin{eqnarray}
\mathcal{E} &=& \eta\frac{v}{g}\partial_i \big(\xi^\dag M_i \xi\big) + \tilde\eta\frac{v}{g^{\prime}}\partial_i\big(f(\rho)G_i\big) \,.
\end{eqnarray}
Thus, we can calculate the mass of the monopole using the Gauss--Ostrogradsky theorem. Taking into account the spherically symmetric Ansatz, we have
\begin{subequations}
\begin{eqnarray}
M &=& 
v \int_{\mathbb{R}^3} \d^3 x \, \partial_i \bigg[\frac{\eta}{g}\xi^\dag M_i \xi + \frac{\tilde{\eta}}{g^\prime}f(\rho)G_i\bigg]
\label{Mintermediate1stline}
\\ &=&
v \lim_{r \rightarrow \infty} \int_{S^2} \d S_i \bigg[\frac{\eta}{g}\xi^\dag M_i \xi + \frac{\tilde{\eta}}{g^\prime}f(\rho)G_i\bigg] 
\\ &=&
v \lim_{r \rightarrow \infty} \int_{S^2} \d \Omega \bigg[\frac{\eta}{g} \underbrace{r x_i \xi^\dag M_i \xi}_{\rightarrow\, - \frac{1}{2} } + \frac{\tilde{\eta}}{g^\prime} f(\rho) \underbrace{r x_i G_i}_{\rightarrow\, 1}\bigg] 
\hspace{1cm}
\\ &=& 4\pi v \bigg[ -\frac{\eta}{2g}  + \frac{\tilde{\eta}}{g^\prime} f(1) \bigg] \,,
\label{Mintermediate}
\end{eqnarray}
\end{subequations}
where we use $\rho(\infty) = 1$ in the last line.

As written, however, this result is ambiguous, for it is not a~priori clear what constant of integration has to be chosen in the primitive function $f$. The answer comes from the condition of applicability of the Gauss--Ostrogradsky theorem: The function, whose divergence is in the integrand, has to be regular everywhere. In our case, however, $G_i$ diverges at the origin; therefore we must set $f(\rho(0)) = f(0) = 0$ in order to make the integrand in Eq.~\eqref{Mintermediate1stline} regular at the origin. We can thus write
\begin{eqnarray}
f(1) &=& \int_0^1 \d \rho \, f^\prime(\rho) \,.
\end{eqnarray}

Furthermore, since we have $\eta = -1$, the first term in the parenthesis in Eq.~\eqref{Mintermediate} is already positive. To make the second term positive too, we must determine $\tilde\eta$ in terms of $f(1)$ as
\begin{eqnarray}
\tilde\eta &=& \sgn f(1) \,.
\end{eqnarray}

To conclude, we can write the final formula for the magnetic monopole mass as
\begin{eqnarray}
\label{eq:massmonopole}
M &=& 
4\pi v \bigg[ \frac{1}{2g} + \frac{1}{g^\prime} \, \Big| \int_0^1\!\d \rho \, f^\prime(\rho) \Big| \bigg]
\,.
\end{eqnarray}
From this result it immediately follows that the lower bound on the mass of the monopole is
\begin{eqnarray}
\label{massbound}
M &\geq&  \frac{2\pi v}{g} \ \approx \ 2.37\,\,\mathrm{TeV} \,.
\end{eqnarray}

The mass formula \eqref{eq:massmonopole} was derived under the assumption of spherical symmetry, but actually, it holds generally. In fact, for a configuration of $n$ monopoles (or $n$ antimonopoles) the formula \eqref{eq:massmonopole} would be just multiplied by $n$.

\subsection{The magnetic charge}

In order to identify the unbroken $U(1)_{\mathrm{em}}$, it is most convenient to switch to the gauge where
\begin{eqnarray}
\xi &=& \left(\begin{array}{c} 0 \\ 1 \end{array}\right) \,.
\end{eqnarray}
This is achieved by the gauge transformation $\xi \to U^\dag \xi$ with
\begin{eqnarray}
U &=& (\tilde \xi, \xi) \,,
\end{eqnarray}
where the charge conjugated field $\tilde\xi \equiv \I \sigma^2 \xi^*$ is also normalized as $\tilde\xi^\dag \tilde\xi = 1$ and, crucially, satisfies $\xi^\dag \tilde\xi = \tilde\xi^\dag \xi = 0$. 
In this gauge, the Lagrangian \eqref{eq:model} reads (switching to physically normalized gauge fields $A_\mu \to g A_\mu$ and $B_\mu \to g^\prime B_\mu$)
\begin{eqnarray}
{\mathcal L}_{\mathrm{eff}} 
&\to& 
\nonumber \\ && \hspace{-8mm} {}
-\frac{1}{4\rho^2} \Big[ \big(F_{\mu\nu}^1\big)^2 + \big(F_{\mu\nu}^2\big)^2 \Big]
+ \frac{v^2 g^2 \rho^2}{8} \Big[ \big(A_\mu^1\big)^2
+ \big(A_\mu^2\big)^2 \Big]
\nonumber \\ && \hspace{-8mm} {}
- \frac{1}{4}f^{\prime 2}(\rho) \big(B_{\mu\nu}\big)^2
+ \frac{v^2\rho^2}{8} \big(g\, A_\mu^3-g^\prime B_\mu\big)^2
\nonumber \\ && \hspace{-8mm} {}
+\frac{v^2}{2}(\partial_\mu\rho) (\partial^\mu\rho)
-\frac{\lambda v^4}{8}\big(\rho^2-1\big)^2
\,.
\end{eqnarray}
Notice that $A_\mu^3$ is not dynamical in our model and should be integrated out. At the leading order, this would give us $A_\mu^3 = g^\prime B_\mu /g$. This means that the standard mass-matrix eigenstate gauge fields are given as
\begin{subequations}
\label{eq:masseigen}
\begin{eqnarray}
A_\mu^{\mathrm{em}} & = & \frac{1}{\sqrt{g^2 +g^{\prime\, 2}}}\Bigl(g^\prime A_\mu^3+g B_\mu \Bigr) = \frac{g^\prime}{e}B_\mu\,, \\
Z_\mu & = &  \frac{1}{\sqrt{g^2 +g^{\prime\, 2}}}\Bigl(g A_\mu^3-g^\prime B_\mu \Bigr) = 0\,,
\end{eqnarray}
\end{subequations}
where $e = gg^\prime/\sqrt{g^2+g^{\prime 2}}$ is the electric charge.

Taking into account now our spherically symmetrical Ansatz, we obtain the magnetic field simply as
\begin{eqnarray}
B_i^{\mathrm{em}} &=& \frac{g^\prime}{e} G_i \ = \ \frac{1}{e} \frac{x_i}{r^3} \,.
\end{eqnarray}
The magnetic charge is thus easily calculated as
\begin{eqnarray}\label{eq:charge}
q &=& \lim_{r \rightarrow \infty} \int_{S^2} \d S_i \, B_i^{\mathrm{em}} 
\ = \ 
\frac{4\pi}{e} \,,
\end{eqnarray}
i.e., twice as big as the Dirac magnetic charge, in accordance with \cite{Cho:1996qd}.

The antimonopole ($q = - 4\pi/e$) is obtained simply by taking instead of $\xi$ its charge conjugation, i.e., instead of Eq.~\eqref{eq:formxi}, by considering the Ansatz
\begin{eqnarray}
\xi &=& \I \left(\begin{array}{c} \cos(\theta/2) \\ \sin(\theta/2) \, \e^{\I\varphi} \end{array}\right) \,.
\end{eqnarray}
The reason is that if $B_i = 2\I\xi^\dag D_i \xi$, then $2\I\tilde\xi^\dag D_i \tilde\xi = -B_i$. 

\begin{center}
*
\end{center}

Let us, in the following, investigate explicit solutions for some particular choices of $f^\prime(\rho)$ (all of them normalized such that $f^\prime(0) = 0$ and $f^\prime(1) = 1$).

\subsection{Power function $f^{\prime}(\rho) = \rho^n$}

First, let us consider a class of functions $f^{\prime}(\rho) = \rho^n$. In order to maintain $f^{\prime}(0) = 0$ we must take $n>0$. Thus, since $f(1) = \int_0^1 \d\rho\, f^\prime(\rho) = 1/(n+1) > 0$, we have $\tilde\eta = 1$. The radial function $\rho$ and the form factor $K$ then come out as
\begin{subequations}
\label{resultspower}
\begin{eqnarray}
\rho(r) &=& \bigg(1+\frac{n-1}{\mu r}\bigg)^{-\frac{1}{n-1}} \,,
\\
\label{powernK}
K(r) &=& \exp\Bigg\{-\frac{g}{g^{\prime}}\frac{\mu r}{2}\bigg(1+\frac{n-1}{\mu r}\bigg)^{-\frac{2}{n-1}}
\\ && \nonumber \hspace{1cm} \times 
\bigg[1- \frac{2}{n+1} \, _2F_1\Big(1,1,\frac{2n}{n-1};-\frac{\mu r}{n-1}\Big)\bigg]\Bigg\}\,,
\end{eqnarray}
\end{subequations}
where $_2F_1(a,b,c,z)$ is the hypergeometric function and where we define for convenience the scale
\begin{eqnarray}
\label{mu}
\mu &\, \equiv \, & v g^{\prime} \, \approx \, 86.1 \,\,\mathrm{GeV} \,.
\end{eqnarray}
Notice that in order to satisfy $\rho(0) = 0$, we must actually take
\begin{eqnarray}
n &\geq& 1 \,.
\end{eqnarray}
The mass of the monopole comes out as
\begin{eqnarray}
M &=& 4\pi v \bigg(\frac{1}{2g} + \frac{1}{(n+1) g^\prime}\bigg) \,.
\end{eqnarray}

It is worth considering explicitly several special values of $n$. Let us first look at the simplest case of a linear function. The general solutions \eqref{resultspower} reduce for $n=1$ to
\begin{subequations}
\label{resultslinear}
\begin{eqnarray}
\rho(r) &=& \exp\bigg\{-\frac{1}{\mu r}\bigg\}
\,,
\\
K(r) &=& \exp\bigg\{ -\frac{g}{g^\prime} \bigg[\frac{\mu r}{2}\exp\bigg(\!\!-\frac{2}{\mu r}\bigg) + \mathrm{Ei}\bigg(\!\!-\frac{2}{\mu r}\bigg)\bigg] \bigg\} \,,
\nonumber \\ &&
\end{eqnarray}
\end{subequations}
where 
\begin{eqnarray}
\mbox{Ei}(x) &=& -\mbox{P}\int\limits_{-x}^{\infty}\frac{\e^{-t}}{t}\d t
\end{eqnarray}
($\mathrm{P}$ stands for the principal value) is the exponential integral with asymptotic behavior $\mathrm{Ei}(x \rightarrow 0) = \log|x| + \gamma_{\mathrm{E}} + \mathcal{O}(x)$.

Despite the presence of the hypergeometric function in Eq.~\eqref{powernK}, the form factor $K(r)$ can in some cases be expressed in terms of elementary functions. To illustrate that, let us present solutions for $n=2$:
\begin{subequations}
\begin{eqnarray}
\rho(r) &=& \frac{1}{\displaystyle 1+\frac{1}{\mu r}}
\,,
\\
K(r) &=& \exp\bigg\{\!\! -\frac{g}{g^\prime} \bigg[ \frac{\mu r}{2} \, \frac{2+\mu r}{1+\mu r} - \log\big(1+\mu r\big)  \bigg] \bigg\} \,,
\hspace{8mm}
\end{eqnarray}
\end{subequations}
and for $n=3$:
\begin{subequations}
\begin{eqnarray}
\rho(r) &=& \frac{1}{\sqrt{\displaystyle 1+\frac{2}{\mu r}}}
\,,
\\
K(r) &=& \exp\bigg\{\!\! -\frac{g}{g^\prime} \bigg[\frac{\mu r}{2} - \log\bigg(1 + \frac{\mu r}{2}\bigg)  \bigg] \bigg\} \,.
\hspace{8mm}
\end{eqnarray}
\end{subequations}

Finally, in the limit $n \to \infty$ the solutions converge to 
\begin{subequations}
\label{resultsninfty}
\begin{eqnarray}
\rho(r) &\xrightarrow[n\to\infty]{}&  1 \,, 
\\
K(r) &\xrightarrow[n\to\infty]{}& \exp \bigg\{-\frac{g}{g^{\prime}} \frac{\mu r}{2}\bigg\} \,,
\end{eqnarray}
\end{subequations}
the energy density goes to
\begin{eqnarray}
\mathcal{E}(r) &\xrightarrow[n\to\infty]{}& \frac{\mu^2}{2g^{\prime 2} r^2} \exp \bigg\{-\frac{g}{g^{\prime}} \mu r\bigg\}
\end{eqnarray}
and the mass saturates the lower bound \eqref{massbound}:
\begin{eqnarray}
M &\xrightarrow[n\to\infty]{}& \frac{2\pi v}{g} \,.
\end{eqnarray}

\begin{figure}[t]
\begin{center}
\resizebox{1.1\columnwidth}{!}{\input{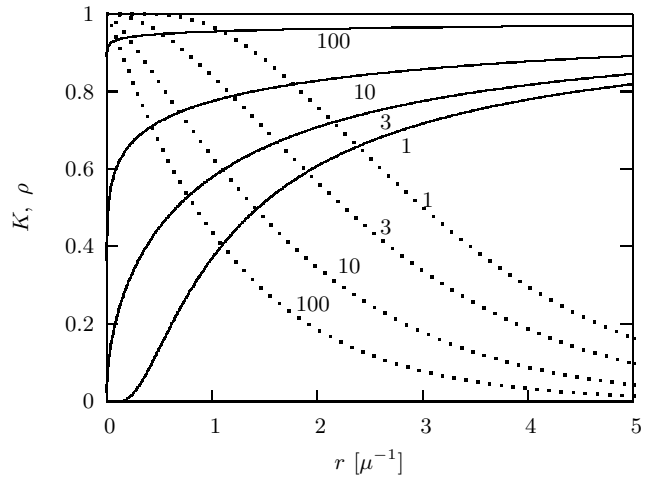}}
\end{center}
\caption{Profiles of $\rho(r)$ (solid lines) and of $K(r)$ (dotted lines) in the case $f^\prime(\rho) = \rho^n$ (see Eq.~\eqref{resultspower}) plotted for various values of $n$.}
\label{fig_power_Krho}
\end{figure}

\begin{figure}[t]
\begin{center}
\resizebox{1.1\columnwidth}{!}{\input{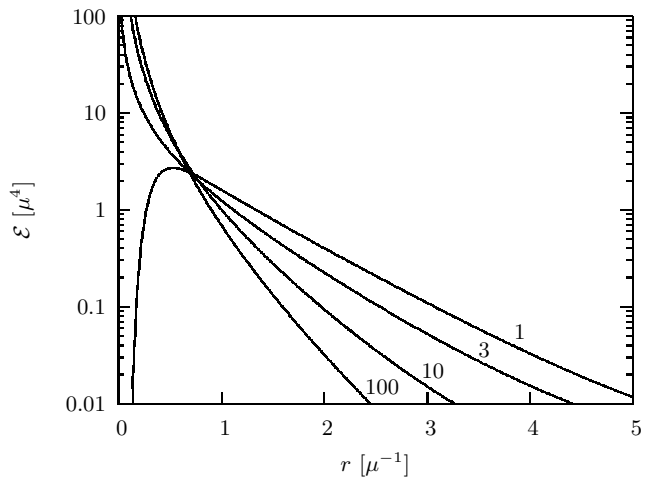}}
\end{center}
\caption{Energy densities ${\mathcal E}(r)$ of the monopole corresponding to the case $f^\prime(\rho) = \rho^n$ for various values of $n$.}
\label{fig_power_E}
\end{figure}

We plot the resulting $\rho$ and $K$ and the corresponding energy densities $\mathcal{E}$ in Figs.~\ref{fig_power_Krho} and \ref{fig_power_E}, respectively.

\subsection{Exponential function $f^{\prime}(\rho) = (\e^{\lambda \rho} - 1)/(\e^{\lambda} - 1)$}

As an example of a transcendental function let us consider the exponential function
\begin{eqnarray}
\label{expfunction}
f^{\prime}(\rho) &=& \frac{\e^{\lambda \rho} - 1}{\e^{\lambda} - 1} \,,
\end{eqnarray}
where $\lambda$ is an arbitrary parameter. Notice that $f^{\prime}(\rho)$ is already normalized to satisfy $f^{\prime}(0)=0$ and $f^{\prime}(1)=1$. For $f(1) = \int_0^1 \d\rho\, f^\prime(\rho)$ we have
\begin{eqnarray}
f(1) &=& \frac{1}{\lambda} -\frac{1}{\e^{\lambda} - 1} \,,
\end{eqnarray}
which is positive for all $\lambda$ (it interpolates between $1$ for $\lambda \to -\infty$ and $0$ for $\lambda \to \infty$), so we set $\tilde\eta = 1$.

\hyphenation{a-na-ly-ti-cal-ly}

The radial function $\rho$ is
\begin{eqnarray}
\rho(r) &=& -\frac{1}{\lambda} \log\bigg[1+\big(\e^{-\lambda}-1\big)\exp\bigg(\! -\frac{\lambda}{\e^{\lambda}-1} \frac{1}{\mu r}\bigg)\bigg] \,,
\nonumber \\ &&
\end{eqnarray}
where we again employ the scale $\mu$; Eq.~\eqref{mu}. This time, however, the form factor $K$ cannot be calculated analytically for general $\lambda$. The mass of the monopole is given as
\begin{eqnarray}
M &=& 4\pi v \bigg[\frac{1}{2g} + \frac{1}{g^\prime} \bigg( \frac{1}{\lambda} -\frac{1}{\e^{\lambda} - 1} \bigg)\bigg]
\end{eqnarray}
and converges to the lower bound $M=2\pi v/g$ for $\lambda \to \infty$.

It is interesting to consider several special limits of $\lambda$, for which $K$ can be computed analytically. Let us first take the limit $\lambda \to -\infty$, in which case the solutions converge to
\begin{subequations}
\begin{eqnarray}
\rho(r) &\xrightarrow[\lambda\to-\infty]{}&  \theta(\mu r-1) \bigg(1-\frac{1}{\mu r}\bigg) \,,
\\
K(r) &\xrightarrow[\lambda\to-\infty]{}&  \theta(1-\mu r) 
\\ \nonumber && {} \hspace{-12mm}
+ \theta(\mu r-1) \exp \bigg\{-\frac{g}{g^{\prime}} \frac{1}{2} \bigg(\mu r - \frac{1}{\mu r} -2 \log \mu r\bigg)\bigg\}  \,,
\end{eqnarray}
\end{subequations}
where $\theta$ is the Heaviside step function.

In the limit $\lambda \to 0$ we obtain the linear function $f^{\prime}(\rho) = \rho$, so the solutions $\rho(r)$, $K(r)$ are the same as Eq.~\eqref{resultslinear} for a power function with $n=1$.

Finally, the limit $\lambda \to \infty$ corresponds to the limit $n \to \infty$ of the power function, so in this limit the solutions are given by Eq.~\eqref{resultsninfty}.

\begin{figure}[t]
\begin{center}
\resizebox{1.1\columnwidth}{!}{\input{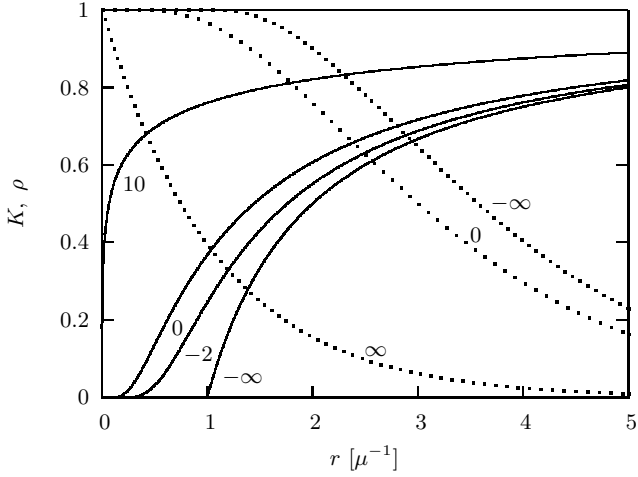}}
\end{center}
\caption{Profiles of $\rho(r)$ (solid lines) and of $K(r)$ (dotted lines) for the exponential function \eqref{expfunction} plotted for various values of $\lambda$.
}
\label{fig_exp_Krho}
\end{figure}

We plot the resulting $\rho$ and $K$ for various $\lambda$ in Fig.~\ref{fig_exp_Krho}.

\section{Generalization}
\label{sec:IV}

The Ansatz for the most general BPS equation reads
\begin{eqnarray}
D_i H &=& \frac{1}{\rho} f_1\, M_i \xi + f_2\, \big(\xi^\dag M_i\,\xi\big)\xi + f_3\, G_i \xi \,,
\end{eqnarray}
where $f_1$, $f_2$, $f_3$ are some functions of $\sqrt{2}|H|/v = \rho$. (The factor $1/\rho$ in the first term is just for convenience.) Notice that no additional gauge covariant terms, linear in $M_i$ and $G_i$ and independent of those already included, can be constructed. 

Being a BPS equation operationally means that the energy density has (in the limit $\lambda \to 0$) a form
\begin{eqnarray}
\mathcal{E} &=& \bigg|D_i H - \frac{1}{\rho} f_1\, M_i \xi - f_2\, \big(\xi^\dag M_i\,\xi\big)\xi - f_3\, G_i \xi \bigg|^2 
\nonumber \\ && {} 
+ \partial_i X_i
\end{eqnarray}
for some $X_i$. In order for this structure to appear,  it turns out that $f_1$ and $f_2$ must be related to each other as
\begin{eqnarray}
f_2 &=& \frac{1}{2} f_1^\prime - \frac{1}{\rho} f_1 
\end{eqnarray}
and the $X_i$ is given by
\begin{eqnarray}
X_i &=& \frac{v}{\sqrt{2}} f_1\, \big(\xi^\dag M_i\,\xi\big) + \sqrt{2} v F_3 G_i\,,
\end{eqnarray}
where $F_3$ is a primitive function of $f_3$, and $\mathcal{E}$ can be written in a manifestly positive form
\begin{eqnarray}
\mathcal{E} &=&  \bigg|\frac{1}{\rho} f_1 \Big(M_i-\xi^\dag M_i\,\xi\Big) \xi + \frac{1}{2} f_1^\prime \big(\xi^\dag M_i\,\xi\big)\xi + f_3\, G_i \xi \bigg|^2
\nonumber \\ && {} \label{eq:genendens}
+ \big|D_i H\big|^2 \,.
\end{eqnarray}

In order to link with previous sections, we rescale the function $f_1(\rho)$, $f_3(\rho)$ as
\begin{equation}
\label{rescale}
f_1 = \eta \frac{\sqrt{2}}{g} h \,,
\hspace{5mm}
f_3 = \tilde\eta \frac{1}{\sqrt{2}g^\prime} f^\prime \,,
\end{equation}
where $f^\prime(\rho)$ is the function already introduced in Sec.~\ref{sec:II} and $h(\rho)$ is a new function. Using these definitions, we see that if the BPS equations
\begin{eqnarray}
\label{BPSgeneral}
D_i H &=&
\eta\frac{\sqrt{2}}{g\rho} h(\rho)\, \Big(M_i -\xi^{\dagger} M_i  \xi\Big)\xi
+ \eta\frac{1}{\sqrt{2}g} h^\prime(\rho)\, \Big(\xi^{\dagger} M_i  \xi\Big)\xi
\nonumber \\ && {}
 +\tilde\eta\frac{1}{\sqrt{2}g^{\prime}}f^{\prime}(\rho)\,G_i\, \xi
\end{eqnarray}
are satisfied, then the Bogomol'nyi bound
\begin{eqnarray}
{\mathcal E} &\geq&
\eta\frac{v}{g}\partial_i \Big(h(\rho)\, \xi^{\dagger} M_i \xi\Big)  + \tilde\eta\frac{v}{g^{\prime}}\partial_i\Big(f(\rho)\,G_i\Big) 
\end{eqnarray}
is saturated. The corresponding Lagrangian\footnote{In principle, it is also possible to modify the kinetic term of the Higgs fields, giving the most general BPS model. Such modifications were considered, e.g., in \cite{Ramadhan:2015qku} (and in references therein). We do not consider such modification here, however, preserving the physical argument that only gauge couplings are to depend on $|H|$. Let us stress, however, that such modifications would not give us more general BPS equations than those discussed here.} possessing a BPS limit (and reducing to the original Lagrangian \eqref{eq:model} of Sec.~\ref{sec:II} for the special case $h(\rho) = 1$) reads
\begin{eqnarray}
\label{eq:model2}
{\mathcal L}_{\mathrm{eff}}
&=& 
\nonumber \\ && {} \hspace{-5mm}
-\frac{v^2}{4 g^2 |H|^2} h^2 \bigg\{ \Tr\big[(F_{\mu\nu})^2\big] - \frac{2}{|H|^4} \Tr\big[F_{\mu\nu} H H^{\dagger}\big]^2
 \bigg\}
\nonumber \\ && {} \hspace{-5mm}
-\frac{1}{4} \bigg(\eta\frac{h^\prime}{g|H|^2} \Tr\big[F_{\mu\nu} H H^{\dagger}\big] + \tilde\eta\frac{f^\prime}{g^\prime} B_{\mu\nu}\bigg)^2
\nonumber \\ && {} \hspace{-5mm}
+|D_\mu H|^2-\frac{\lambda}{2}\bigg(H^{\dagger}H -\frac{v^2}{2}\bigg)^2
\,.
\end{eqnarray}
As with the function $f^\prime$ (cf.~Eq.~\eqref{fprimein1}), we require the normalization
\begin{eqnarray}
\label{hin1}
h^2(1) &=& 1\,.
\end{eqnarray}
If the above condition is not met, we can always absorb $h^2(1) \not = 0$ into the definition of $g^\prime$. In other words, we can demand Eq.~\eqref{hin1} without loss of generality.

Analyzing the generalized BPS equations \eqref{BPSgeneral} in the same way, we find that $B_i$ is given again by Eq.~\eqref{eq:ident}, while Eq.~\eqref{eq:rho1} for $\rho$ and the reduced BPS equations \eqref{eq:bpsred} modify to
\begin{eqnarray}
\partial_i \rho &=& \frac{\eta}{gv} \Big(\xi^{\dagger} M_i \xi\Big) h^\prime + \frac{\tilde\eta}{g^\prime v} G_i f^\prime
\end{eqnarray}
and
\begin{eqnarray}
\label{eq:bpsred2}
\bigl(\unitmatrix{2} -\xi\xi^{\dagger}\bigr)\bigg(D_i \xi - \eta\frac{2}{gv\rho^2} h M_i\xi \bigg) 
&=& 0 \,,
\end{eqnarray}
respectively. These equations reduce for the spherically symmetric Ansatz to
\begin{subequations}
\label{EOMspherical2}
\begin{eqnarray}
\label{eq:rho2}
\rho^\prime &=& \frac{\eta}{gv} \frac{K^2-1}{2r^2} h^\prime + \frac{\tilde\eta}{g^\prime v} \frac{1}{r^2} f^\prime \,,
\\
\label{eq:K2}
\frac{K^{\prime}}{K} &=& \eta\frac{g v}{2} \frac{\rho^2}{h} \,;
\end{eqnarray}
\end{subequations}
cf.~Eq.~\eqref{EOMspherical}. From the second equation is it thus evident that in order to meet the boundary condition $K(\infty) = 1$ we must set
\begin{eqnarray}
\label{etagen}
\eta &=& - \sgn h(1) \,.
\end{eqnarray}
On the other hand, the requirement that the integral of the energy density, rewritten using the equation of motion as
\begin{eqnarray}
\label{eq:enKform2}
\mathcal{E} &=& \frac{v^2}{2}\frac{\rho^2K^2}{r^2}  + \bigg(\frac{\eta}{g} \frac{K^2-1}{2r^2} h^{\prime} + \frac{\tilde\eta}{g^{\prime}} \frac{1}{r^2}f^{\prime} \bigg)^2 \,,
\end{eqnarray}
converges at the lower limit does not give us any constraint on the value of $h^\prime$ at $r=0$ due to $K(0) = 1$.

Using the Gauss--Ostrogradsky theorem we can calculate the mass of the monopole as
\begin{eqnarray}
M &=& 4\pi v \bigg[ -\frac{\eta}{2g} h(1)  + \frac{\tilde{\eta}}{g^\prime} f(1) \bigg] \,,
\end{eqnarray}
which, curiously, yields the \emph{same} expression as in Eq.~\eqref{eq:massmonopole} upon taking into account Eqs.~\eqref{hin1} and \eqref{etagen}. In particular, the lower bound \eqref{massbound} is unaltered!

Finally, let us  consider the Higgs vacuum $\rho \to 1$ and the gauge $\xi = \left(\begin{smallmatrix}0 \\ 1\end{smallmatrix}\right)$. Let us also switch to physically normalized gauge fields, i.e. $A_\mu \to g A_\mu$ and $B_\mu \to g^\prime B_\mu$.
The above Lagrangian becomes
\begin{eqnarray}
{\mathcal L}_{\mathrm{eff}} &\to&
- \frac{1}{4} \bigg[ \big(F_{\mu\nu}^1\big)^2 + \big(F_{\mu\nu}^2\big)^2 \bigg]
+ \frac{g^2 v^2}{8} \bigg[ \big(A_\mu^1\big)^2 + \big(A_\mu^2\big)^2 \bigg]
\nonumber \\ && {}
- \frac{1}{4} \bigg( \frac{\eta}{2} h^\prime(1) F^3_{\mu\nu} - \tilde\eta f^\prime(1) B_{\mu\nu} \bigg)^2
\nonumber \\ && {}
+ \frac{v^2}{8} \big(g\, A_\mu^3-g^\prime B_\mu\big)^2
\,.
\end{eqnarray}
As was the case in Sec.~\ref{sec:II} only a particular combination of $A_\mu^3$ and $B_\mu$ remains dynamical. In the generalized model, however, this is true only in the limit $\rho \to 1$ and for $\rho \not = 1$, all fields are dynamical.

Nevertheless, to identify the electromagnetic field the limit $\rho \to 1$ is sufficient. If we eliminate the non-dynamical degrees of freedom from the above Lagrangian, we obtain at the leading order $A_\mu^3 = g^\prime B_\mu /g$ and the mass-matrix eigenstates would be given exactly the same as in Eqs.~\eqref{eq:masseigen}. Consequently, the magnetic charge will be the same as in Eq.~\eqref{eq:charge}.

\section{Universality of the lower bound}
\label{sec:V}

As we saw in the previous section, the mass of the monopole cannot be lower than $2\pi v/g$ in all BPS theories that are classified via functions $h$ and $f^\prime$. We claim that this bound also applies to any \emph{non-BPS} model that supports the Cho--Maison monopole.

Heuristically, the argument goes as follows. The BPS mass depends only on the asymptotic behavior of fields at the spatial infinity. This behavior should be the same across all theories since it is fixed, among other things, by the underlying topology. (Of course, we assume that the field content, as well as the gauge group, is the same for all models under consideration.) In non-BPS theories, however, the mass also depends on the precise behavior of fields throughout the volume. Intuitively, this contribution should be positive. If so, we obtain a plausible-sounding statement that the mass of a topological soliton in non-BPS theories is larger than the BPS mass. 

However, a possible subtlety lies in the fact that it is not clear whether one can separate the topological contributions from non-topological ones and maintain the positive definitiveness of the latter. It seems, however, very unlikely  that the total non-topological energy can come out negative for an a priori positive energy functional. Nevertheless, it seems hard to prove this assertion without knowledge of the precise structure of the theory at hand.

Let us, therefore, concentrate on a concrete model where we can articulate our claim precisely. As an example, let us look at the non-BPS theory of \cite{Cho:2013vba} and \cite{Ellis:2016glu}, whose energy density we write in our notation as
\begin{eqnarray}
{\mathcal E}[\epsilon] 
&=& \frac{1}{g^2}\Tr\big[M_i^2\big]+\frac{1}{2g^{\prime 2}}\epsilon\, G_i^2
\nonumber \\ && {} 
+ |D_i H |^2 + \frac{\lambda}{2}\bigg(HH^\dagger-\frac{v^2}{2}\bigg)^2 \,,
\end{eqnarray}
where $\epsilon$ is a positive function of $\rho$. In particular, $\epsilon =1$ gives us the bosonic part of the electroweak model. To prove our claim, we have to compare this energy density with the energy density of the general BPS theory ${\mathcal E}[h,f^\prime]$ given in Eq.~\eqref{eq:genendens}. Concretely, we have to show that for a given $\epsilon$ there exist $h$ and $f^\prime$ such that
\begin{eqnarray}
{\mathcal E}[\epsilon] &\geq& {\mathcal E}[h, f^\prime] \,.
\end{eqnarray}
Indeed, we have
\begin{eqnarray}
{\mathcal E}[\epsilon]  - {\mathcal E}[h, f^\prime] 
&=& \frac{1}{g^2}\Tr\big[M_i^2\big]\bigg(1-\frac{h^2}{\rho^2}\bigg)
\nonumber \\ && {} 
+\frac{1}{2g^{\prime\, 2}}G_i^2\big(\epsilon-f^{\prime\, 2}\big)\nonumber
\nonumber \\ && {} 
+\frac{1}{2g^2} \big(\xi^\dag M_i\,\xi \big)^2 \bigg(\frac{4 h^2}{\rho^2}-h^{\prime\, 2}\bigg)
\nonumber \\ && {} 
-\frac{\eta \tilde\eta}{g g^\prime}h^\prime f^\prime G_i \big(\xi^\dag M_i\,\xi \big) \,.
\end{eqnarray}
If we now express the functions $f^\prime$ and $h$ in terms of a new function $\sigma$ as
\begin{subequations}
\label{ansatzfprime2h}
\begin{eqnarray}
f^{\prime\,2} &=& \epsilon-\frac{\epsilon}{4}\bigg(1-\frac{\rho \sigma \sigma^\prime}{1+\sigma^2}\bigg)^2 \,, 
\\
h &=& \frac{\rho}{\sqrt{1+\sigma^2}} \,, 
\end{eqnarray}
\end{subequations}
where $\sigma$ satisfies $\sigma(1)=0$ (so that $h(1) = 1$) and
\begin{eqnarray}
1+\sigma^2 &\geq& -\rho \sigma \sigma^\prime
\hspace{1cm} \mbox{(for all $\rho$)}
\end{eqnarray}
(so that $f^{\prime\,2} \geq 0$), the right-hand side of the above equation takes on a manifestly positive form:
\begin{eqnarray}
{\mathcal E}[\epsilon]  - {\mathcal E}[h, f^\prime] 
&=& \frac{1}{g^2}\Tr\big[M_i^2\big]\frac{\sigma^2}{1+\sigma^2}
+\frac{1}{2}X_i^2 \,,
\end{eqnarray}
where
\begin{eqnarray}
X_i &=&
\frac{\eta}{g} \frac{2}{\sqrt{1+\sigma^2}}
\sqrt{1-\frac{1}{4}\bigg(1-\frac{\rho \sigma \sigma^\prime}{1+\sigma^2}\bigg)^2}
\big(\xi^\dag M_i\,\xi \big)
\nonumber \\ && {}
- \frac{\tilde \eta}{2g^\prime} \sqrt{\epsilon} \, \bigg(1-\frac{\rho \sigma \sigma^\prime}{1+\sigma^2}\bigg)
G_i
\,.
\end{eqnarray}
In other words, we see that ${\mathcal E}[\epsilon] \geq {\mathcal E}[h, f^\prime]$ is true for $h$, $f^\prime$ given by the Ansatz \eqref{ansatzfprime2h}, provided $\sigma$ satisfies the above requirements. The simplest possibility is to take $\sigma = 0$, corresponding to $f^{\prime 2} = 3\epsilon/4$ and $h = \rho$, in which case we obtain
\begin{eqnarray}
{\mathcal E}[\epsilon]  - {\mathcal E}[\rho, \sqrt{3\epsilon}/2] &=&
\frac{1}{2}\bigg(
\frac{\eta}{g}\sqrt{3}\big(\xi^\dag M_i\,\xi \big)
-
\frac{\tilde\eta}{2g^\prime}\sqrt{\epsilon}\,G_i
\bigg)^2 \,.
\nonumber \\ &&
\end{eqnarray}
This shows that the monopoles in the \qm{$\epsilon$ model} cannot be lighter than our universal bound.

\section{Discussion}
\label{sec:VI}

In this paper, we have presented a family of effective modifications to the electroweak model that contains BPS magnetic monopoles as classical solutions. These solutions are the BPS extensions of Cho--Maison monopoles presented in \cite{Cho:2013vba} and further studied in \cite{Ellis:2016glu}. We have also 
studied several explicit examples, where we found monopole solutions in analytic form. 

Critically, we obtained a universal lower bound on the mass of the monopole as $M \geq 2\pi v/ g \approx 2.37\,\,\mathrm{TeV}$. In Sec.~\ref{sec:V} we argued that this bound also applies to non-BPS models, in particular to the model presented in \cite{Cho:2013vba} and \cite{Ellis:2016glu}.

The modifications of the electroweak theory that are novel in our model are the first two terms in Eq.~\eqref{eq:model}. Interestingly, the structure of these modifications turns out to be completely fixed by the requirement that the coupling between the $SU(2)_L$ and $U(1)_Y$ gauge fields disappear. In Sec.~\ref{sec:IV}, we relax this requirement and showcase a general family of BPS theories. Nevertheless, we find that the formula for the BPS mass \eqref{eq:massmonopole} is exactly the same as in the model \eqref{eq:model}.

As we have seen, the lack of the dynamical $Z$ boson in the perturbative spectrum is the price one has to pay for a BPS Cho--Maison monopole. Let us offer some physical reasoning that might explain this rather curious property of BPS models. As Manton showed \cite{Manton:1977er}, the attractive scalar interaction between well separated 't~Hooft--Polyakov monopoles is exactly balanced with the repulsive Coulomb force in the BPS limit. While the BPS models that we discuss here are different from the pure $SU(2)$ model of 't~Hooft and Polyakov, it is telling that the perturbative spectra are the same. It is not unreasonable to think that the presence of an additional massive particle ($Z$ boson) would upset this balance. Although we have not shown in this work that there is a similar cancellation of forces between two Cho--Maison monopoles, we strongly suspect that that is the case. Therefore, we expect that the reason behind the lack of the $Z$ boson in the spectrum is precisely to allow static multi-monopole configurations to exist. We leave the proof of this assertion for future work.

Our work opens up the possibility of studying multi-particle configurations of Cho--Maison monopoles, which are sometimes characterized as conceptually being something between Dirac's monopole and the 't~Hooft--Polyakov monopole. To construct a multi-monopole configuration of Dirac's monopoles is not particularly challenging. On the other hand, to obtain a 't~Hooft--Polyakov multi-monopole configuration requires the use of highly sophisticated tools, such as the Nahm construction \cite{Nahm:1981nb}. It would make an interesting future study to elaborate how difficult is to write down a multi-monopole configuration of Cho--Maison monopoles in the BPS limit and whether one can adopt the Nahm approach or other techniques.

Furthermore, our model can be a useful tool to explore non-topological solutions of the electroweak model, such as the spharelon \cite{Klinkhamer:1984di}. Indeed, the BPS limit lends itself to the possibility that we can find unstable static solutions in the analytic form. We plan to investigate this in the future work.

As far as we know our theory is not a bosonic part of any known supersymmetric theory. It would be interesting to see if there actually exists a SUSY extension of our model and whether the BPS Cho--Maison monopole can be constructed as a 1/2, 1/4, or other SUSY state. We leave this puzzle for the future.

\acknowledgments

This work was supported by the Albert Einstein Centre for Gravitation and Astrophysics financed by the Czech Science Agency Grant No.~14-37086G (F.~B.) and by the program of Czech Ministry of Education, Youth and Sports INTEREXCELLENCE Grant No.~LTT17018 (F.~B., P.~B.). P.~B. also thanks TJ~Balvan Praha for support.


\providecommand{\href}[2]{#2}\begingroup\raggedright\endgroup

\end{document}